\documentstyle[prl,preprint,aps,psfig]{revtex} 

\global\firstfigfalse

\begin{document}

\draft 
\title{Characterization of Quantum Chaos by the Autocorrelation
  Function of Spectral Determinants} 
  \author{S. Kettemann
  \footnote[2]{Present Address: Max-Planck Institut f\"ur Physik
  Komplexe Systeme, Au\ss enstelle Stuttgart, 70569 Stuttgart,
  Germany}, D. Klakow \footnote[3]{Present Address: Institut f\"ur
  Theoretische Physik II, Staudtstr. 7, Univ. Erlangen, 91058
  Erlangen, Germany} and U. Smilansky}
\address{ {Department of  Physics of Complex Systems,} \\
  {The Weizmann Institute of Science, Rehovot 76100, Israel}}
\date{\today} 
\maketitle

\begin{abstract}
  The autocorrelation function of spectral determinants is proposed as
  a convenient tool for the characterization of spectral statistics in
  general, and for the study of the intimate link between quantum
  chaos and random matrix theory, in particular. For this purpose, the
  correlation functions of spectral determinants are evaluated for
  various random matrix ensembles, and are compared with the
  corresponding semiclassical expressions.  The method is demonstrated
  by applying it to the spectra of the quantized Sinai billiards in
  two and three dimensions.
\end{abstract}

\section{Introduction}

One of the most important discoveries in the study of quantum systems
which display chaotic dynamics in the classical limit, was the fact
that generically, the spectral statistics obey the predictions of
Random Matrix Theory (RMT). This observation was originally made on
the basis of numerical experiments (See e.g, \cite {BohigasLH} for
energy spectra, \cite {USLH} for $S$-matrix spectra), but during the
past years it received a much stronger theoretical foundation. (\cite
{BerryA400} \cite {AA} \cite {AAA} \cite {Bogolkeating}). The standard
tools for the quantitative study of spectral statistics were the
n-point correlation functions, or some functions thereof which were
usually chosen because of their suitability in the analysis of finite
spectral stretches. In the present work we would like to propose a
different approach, which is based on the study of the statistical
properties of the {\it spectral determinant}, sometimes referred to as
the {\it secular function}, the {\it spectral $\zeta$ function} or the
{\it characteristic polynomial}: It is the function which vanishes if
and only if its argument belongs to the spectrum. The spectral
determinant, considered as a function of its variable, is defined in
terms of coefficients which can be expressed as functions of the
eigenvalues. This is particularly simple for the case of a matrix of
finite dimension $N$, where the characteristic polynomial is defined
in terms of its coefficients, which, in turn, can be calculated from
the spectrum. Thus, the information stored in the spectral function is
equivalent to the information stored in the spectrum, and one may ask
why should one study the statistics of the former, when the n-point
spectral statistics are so well investigated.

In this article we shall try to show that in fact the correlation
function may offer some important advantages for studying the
statistics of the spectrum.  To name a few: In general the
coefficients in the spectral determinants depend in a complicated way
on the spectral points. Sometimes, however, the ensemble averages of
these quantities have particularly simple expressions, which are
convenient for numerical and theoretical studies. This was the reason
why the Essen group \cite {haake} studied the statistical properties
of the coefficients of the characteristic polynomials for the random
Circular ensembles. We would like to emphasize other aspects, which
are especially important for establishing the connection between
quantum chaos and RMT. The Gutzwiller trace formula, which gives the
semiclassical theory for the spectral density, typically diverges. The
derivation of spectral statistics based on this theory, should
therefore be augmented by additional assumptions \cite {BerryA400} or
ad-hoc truncation procedures \cite {Bogolkeating}. In contrast, the
semiclassical expression for the spectral determinant involves a
finite number of periodic orbits, and therefore it converges on the
real energy axis.  The semiclassical spectral determinant preserves
another important property, namely, it is explicitly real for real
energies (\cite {Bogomolny}, \cite {DoronUS}, \cite {KeatingBerry}).
Thus, the semiclassical study of the statistical properties of the
spectral determinant can be based on a relatively solid starting
point. Last but not least, the semiclassical spectral determinant
shares many of its properties with the Riemann Siegel expression for
the Riemann $\zeta$ function on the critical line. The autocorrelation
function for $\zeta _R$ is the subject of a study of Keating {\it et
al} \cite {Keatingzeta} which was carried out in parallel to the
present work. The rigorous results obtained for the $\zeta_R$ case,
provide some support to the physically reasonable, (yet mathematically
uncontrolled) approximations made in the semiclassical theory to be
discussed here.

The object we would like to study in the present paper is the
autocorrelation function. Denote by $Z(x)$ the secular function, which
vanishes at the spectral points $\{ x_n \}$. The variable $x$ can
stand for the energy, or for the phase variable when the operator
under study is a unitary operator such as a $S$ matrix or a unitary
evolution operator. Consider a domain of size $\Delta$ centered about
$x_0$. The autocorrelation function is defined in terms of

\begin{equation}
{\bar C} (\xi;x_0) = {1 \over \Delta}
\int_{x_0-\Delta/2}^{x_0+\Delta/2} Z(x+\xi/2) Z^*(x-\xi/2) {\rm d}x .
\label {corr1}
\end{equation}
The mean, normalized autocorrelation function is
\begin {equation}
  C(\xi) = {< {\bar C}(\xi;x_0)> \over <{\bar C} ( 0;x_0) >} \label
  {corr}
\end {equation}
The brackets stand for averaging over an appropriate random matrix
ensemble, or for a spectral average which is affected by averaging
over $x_0$. It is assumed throughout that the correlation variable
$\xi$ takes values which are always much smaller than the integration
interval $\Delta$.

The behaviour of $C(\xi)$ can be intuitively clarified, by considering
two extreme cases: An equally spaced (infinitely rigid) spectrum
produces a correlation function which is a strictly periodic function
of $\xi$,
\begin {equation}
  C(\xi) = \cos \pi \xi
\end{equation}
Where $\xi$ is measured in units of the level spacing. On the other
hand, a Poissonian spectrum with $N$ spectral points, yields a
positive correlation which decays to zero on a scale which is
proportional to ${\sqrt N}$.  This can be shown by considering the
ensemble of diagonal matrices with independent, Gaussian random
diagonal elements with $< H_{ii}^2 > = {N^2 \over 2\pi} $.  With this
choice, the average level spacing at $E=0$ is $1$.  The correlation
function is
\begin{equation}
C_N(\xi)= ( 1 - {\xi^2\over 2\pi N^2} )^N \rightarrow \exp ( -
{\xi^2\over 2\pi N})
\end{equation}

Thus, the lack of correlation between the energy levels, induces a
slowly decaying correlation function.

The canonical random ensembles all display level repulsion which
induce strong correlations.  It is expected, therefore, that the
"incipient crystalline character" \cite {Dyson} of the spectrum of
these canonical random matrix ensembles will manifest itself by
oscillatory, yet slowly decaying correlation functions. The more rigid
the spectrum (larger $\beta$), the more marked and persistent will be
the oscillations of $C(\xi)$.

The paper is organized in the following way. The next chapter will be
devoted to the derivation of the autocorrelation function for the
standard random matrix ensembles.  Next, we shall discuss the
semiclassical derivation, and compare it to the prediction of RMT.
Finally, we shall present some numerical results which were performed
to illustrate and to supplement the theoretical derivations.

 
\section {Random Matrix Theory for the Autocorrelation Functions} 
\label{theory}

The autocorrelation function takes a particularly simple form in the
case of the circular random ensembles.  Moreover, the recent results
of the Essen group \cite {haake} can be directly used to obtain closed
expressions for the autocorrelation function. For the sake of
completeness, we shall briefly summarize the results for the circular
ensembles in the first part of this section. The Gaussian ensembles
require some more work: correlation functions of the kind we are
interested in, were only derived for the unitary (GUE) ensemble \cite
{AS}. Moreover, the semicircular mean spectral density for the
Gaussian ensembles might introduce irrelevant features in the
correlation functions, if they are not properly treated. The
derivation of the autocorrelation function for any Gaussian ensemble
which interpolates between the GOE and the GUE, will be given in the
second part of this section, together with a discussion of the effects
which are due to the non uniform mean spectral density.  Fortunately,
we find that these effects become less significant as the dimension of
the random matrices increases.

\subsection {The Circular Ensembles}

The spectrum of a $N \times N$ unitary matrix $S$ consists of $N$
unimodular eigenvalues $e^{i\theta_l}, \ \ 1\leq l \leq N $.  It is
convenient to write the characteristic polynomial such that it is real
on the unit circle
\begin{equation}
Z_S(\omega)= e^{{i\over2}(N\omega - \Theta)} \det (I-e^{-i\omega}S)
\end{equation} 
Where $e^{i\Theta} = \det (-S) $. The characteristic polynom can be
written down as
\begin{equation}
Z_S(\omega)= e^{{i\over2}(N\omega - \Theta)} \sum_{l=0}^N
a_le^{-i\omega l}
\end{equation}
The autocorrelation function reads now
\begin{equation}
{\bar C} (\xi) = {1 \over 2\pi} \int_{0}^{2\pi}
Z_S(\omega+\xi/2)Z_S(\omega-\xi/2){\rm d}\omega
\label {corrS1}
\end{equation}
or explicitly,
\begin {equation}
  C_{\beta}(\xi) = {\sum_{l=0}^N <|a_l|^2>_{\beta} e^{i \xi (l-{N\over
        2})} \over \sum_{n=0}^N <|a_n|^2>_{\beta}}
\label {corrcen}
\end{equation}
where $< \cdot >_{\beta} $ stands for the average with respect to the
spectral measure of the orthogonal $(\beta =1)$, unitary $(\beta =2)$
or symplectic $(\beta=4)$ circular ensembles.  The unitarity imposes
the identity $|a_n| = |a_{N-n}|$ which guarantees that the
autocorrelation function is real, and reduces the effort needed to
calculate the ensemble averages.

The ensemble averages $< |a_n|^2 >_{\beta} $ were calculated in \cite
{haake} for all values of $\beta$. We shall quote here the results for
the orthogonal and the unitary ensembles:

\begin{equation}
<|a_n|^2>_{\beta =1} =1 +{n(N-n) \over N+1} \ \ ; \ \
<|a_n|^2>_{\beta =2} = 1.   \ \ \ \  {\rm for}\ \ n=0 \dots N
\label {haake}
\end {equation}
It is convenient to introduce the scaled correlation length $\omega= \xi
{2\pi\over N}$ and in
the limit of large $N$ the sums in (\ref {corrcen}) can be approximated as
integrals, to give

\begin {equation}
C_{\beta = 1}(\omega) = {3\over 2} \left ( {\sin\pi \omega
\over \pi \omega} + {1\over \pi^2}
{\partial ^2\over \partial \omega^2}{\sin \pi \omega \over \pi \omega}\right) \
 \
\ ;  \ \ \
C_{\beta = 2}(\omega) = {\sin\pi \omega \over \pi \omega }
\label {corrcircle}
\end {equation}

Equation (\ref {corrcircle}) suggests the following relations
between the real correlation functions:
\begin{equation}
C_{\beta=1}(\omega) = -3 {1 \over \pi^2 \omega }{\partial
\over\partial \omega} C_{\beta=2}(\omega)
\label {crelation}
\end{equation}

The two correlation functions are displayed in Fig.~\ref{exgaus}.  As
expected, the more rigid ensemble ($\beta =2$) shows stronger
correlations than the less rigid ensemble.

The expressions derived above for the circular ensembles appear also
when the Gaussian ensembles are discussed in the large N limit. A
detailed derivation will follow in the next subsection.

\subsection {The Gaussian Ensembles}

Unlike the circular ensembles, the mean spectral densities for the
Gaussian ensembles are not uniform. This has to be incorporated in the
definition of the correlation function for the Gaussian ensembles.
Otherwise this spurious behavior may blur the essential
correlations. Because of this consideration we shall compute the
functions

\begin{equation}
\bar{C}_N(E,\omega) = \left< \det ( E +{\omega \over 2} - H )
\det (E - { \omega  \over 2} - H) \right>_H.
\end{equation}
 
 This function should then be normalized by its value at $\omega =0$,
 to get the function $C(\omega)$, as defined in Eq. ( \ref{corr} ) If
 we will find a dependence of this function on E, even for large N, we
 will have to perform an integration over a restricted region of the
 center E: $|E| \leq {\sqrt { N}}$.  $<\cdot >_H$ indicates an average
 over the ensembles of hermitian matrices with respect to the measures
 which will be defined below.

 Before doing any calculations, one can get some insight in this
  function by writing $C_N$ as
\begin{eqnarray}
\bar{C}_N(E,\omega) &=& \left< \prod_{i=1}^{N}( E- { \omega  \over 2}- E_i)
(E+{ \omega \over 2} -E_i)\right>_H \nonumber \\ &=& \left<
 \exp(\sum_{i=1}^N (\ln(E-{ \omega \over 2} -E_i) + \ln(E + { \omega
 \over 2} - E_i)) \right>_H.
\end{eqnarray}
  Hence, $C_N$ can be considered as the partition function of a gas of
particles at positions $E_i$, moving in a quadratic potential (which
is provided by the Gaussian measures) and interacting via a
logarithmically repulsive interaction with two particles whose
positions are fixed at $E +{ \omega \over 2}$ and $E- {\omega \over
2}$ and which do not interact with each other.  The interaction
between the particles at positions $E_i$ themselves depends on the
distribution and the symmetry of the random hamiltonian $H$ as we will
see in more detail below.  It may be useful to come back to this
Coulomb gas picture of $C_N$ to understand the results whose
derivation we present in the following.

 For analytical calculations within an ensemble of Random matrices, it
is more convenient to write each determinant as the functional
integral over two sets of N Grassmann variables, which greatly
simplifies the averaging over the weight of the ensemble:
\begin{equation}
\bar{C}_N(E,\omega) = \int d \xi d \xi^* d\eta d\eta^*\left< e^{-\xi^+
 ( E +{ \omega \over 2} - H) \xi} e^{-\eta^+ ( E - { \omega \over 2} -
H ) \eta }\right>_H \label{g1},
\end{equation}
 where $\xi^T= (\xi_1,...,\xi_N)$, $\eta^T=(\eta_1,...,\eta_N)$,
$\xi^+=(\xi^*_1,...,\xi^*_N)$, $\eta^+=(\eta^*_1,...,\eta^*_N)$, and
with
\begin{equation}
\int d\xi d\xi^* e^{-\xi^* A \xi} = det A,
\label{g2}
\end{equation}
\begin{equation}
\int d\xi d\xi^* \xi^*_i \xi_j e^{-\xi^* A \xi} = (A^{-1})_{ij} \det A.
\end{equation}
 Now, the averaging over the random matrices $H$ is reduced to a
 simple Gaussian integral, when they are Gaussian distributed.  Then,
 performing the Grassmann integrals we could find $C_N(E,\omega)$ as
 an int( N/2 )-th order polynomial in $\omega^2$, and we could compare
 its coefficients with the ones obtained numerically for the non
 integrable quantum systems, for arbitrary N.  Although we succeeded
 to do this for arbitrary N, it turns out to be more useful to do a
 saddle point approximation which coincides with the exact result in
 the limit $N \rightarrow \infty$.

 In the present section we shall present a theory which treats the
unitary and the orthogonal Gaussian random ensembles (and any
interpolating ensemble) in a unified way. This is done by writing the
Hamiltonian as\cite{altland}
\begin{equation}
H = H_s + i \alpha H_a, \label{m1}
\end{equation}
where $H_s$ is a symmetric real matrix, its matrix elements $H_{s ij}=
 H_{s ji}$ are real numbers $ \forall i, j=1,...,N $.  The matrix
 elements of $H_a$ are antisymmetric, $H_{a ij} = - H_{a ji}$ and real
 numbers $ \forall i\neq j =1,...,N$, since the Hamiltonian H has to
 be hermitian.

The random matrices are distributed with a Gaussian measure,
\begin{eqnarray}
&& \exp ( - c {\rm Tr} ( H^2 ) ) {\rm d}H \nonumber \\ &=& \exp ( - c
 \sum_{i=1}^N H_{s i}^2 - 2 c
\sum_{i > j} H_{s ij}^2 - 2 c \alpha^2
\sum_{i > j} H_{a ij}^2 ){\rm d}H.
 \label{m2}
\end{eqnarray}
where $c=\pi^2/(4N)$ corresponds to a spectrum with mean level spacing
set equal to $d=1$.

Note that, $ \alpha=0$ corresponds to GOE where there is time reversal
 symmetry, the Hamiltonian $H$ is symmetric and the distribution is
 invariant under orthogonal transformations $H \rightarrow O H
 O^{-1}$.  $ \alpha=1$ describes the pure GUE where the time reversal
 symmetry is broken completely and the Hamiltonian is hermitian and
 invariant under unitary transformations $H \rightarrow U H U^{-1}$.

 Calculating the correlation function for $\alpha$ between 0 and 1, we
 obtain the crossover between GOE and GUE.

Using Eq. (\ref {g1}) we arrive after performing the Gaussian average
over the random matrices at:
\begin{eqnarray}
\bar{C}_N(E,\omega) &=& \int d \xi d \xi^* d \eta d \eta^* \exp ( i\sum_{i=1}^N
\xi_i^+ (E + { \omega \over 2}) \xi_i
+ i \sum_{i=1}^N \eta_i^+ (E - { \omega \over 2}) \eta_i ) \nonumber
\\ && \exp ( - \frac{1}{4 c} \sum_{i=1}^N (\xi^*_i \xi_i + \eta^*_i
\eta_i)^2)
\exp \left(- \frac{1}{8 c} \sum_{i > j} (\xi^*_i \xi_j + \xi_j^* \xi_i
+ \eta_i^*
\eta_j + \eta_j^* \eta_i)^2 \right) \nonumber \\
&&
\exp \left( \alpha^2 \frac{1}{8 c} \sum_{i>j} (\xi^*_i \xi_j - \xi^*_j \xi_i
+\eta^*_i \eta_j -\eta^*_j \eta_i)^2 \right).
\label{p1}
\end{eqnarray}
  Thus, we arrived at an interacting theory with interaction strength
$1/(4c)$.

 Although the Grassmannian functional integrals can be performed
exactly, let us treat this expression as one usually does when one
encounters an interacting theory: First do a Hubbard- Stratonovitch
transformation. If the resulting decoupled theory can not be solved
exactly, do a saddle point approximation. Expansion around the saddle
point then leads to a nonlinear sigma model theory which can possibly
be integrated out.  The advantage over the exact result turns out to
be that the limit $N \rightarrow \infty$ can be done more easily and
is exact.  We may first write Eq. (\ref{p1}) more compactly (compare
with Refs.
\onlinecite{altland},\onlinecite{AS}):
\begin{eqnarray}
\bar{C}_N(E,\omega) &=& \int d \psi \exp (\frac{1}{2} i \sum_i
\bar{\psi}_i (E + { \omega \over 2} \Lambda ) \psi_i )
\exp \left( - \frac{1}{16 c} {\rm Tr} ( \sum_{i=1}^N \psi_i \otimes
\bar{\psi}_i)^2
\right) \nonumber \\
&&\exp \left(- \frac{\alpha^2}{16 c} {\rm Tr} ( \sum_{i=1}^N \psi_i
 \otimes \bar{\psi}_i
\tau_3 )^2 \right),
\end{eqnarray}
where
\begin{equation}
\Lambda= \left( \begin{array}{cc} \openone_{2 x 2} & 0 \\
                                   0              & - \openone_{2 x 2}
                 \end{array} \right),
\end{equation}
ensures that the $\xi's$ get a $ ``+ \omega''$ and the $\eta's$ get a
$''-\omega''$.
\begin{equation}
 \tau_3 = \left( \begin{array}{cc} 1 & 0 \\ 0 & -1 \end{array}
\right),
\end{equation}
has to be introduced due to the antisymmetric part of the random
matrices $H_a$.  For compactness the vectors of anticommuting
variables were introduced:
\begin{equation}\label{psi}
 \psi_i = \left( \begin{array}{c} \xi_i \\ \xi_i^* \\
\eta_i \\ \eta_i^* \end{array} \right),
 \bar{\psi}_i = ( \xi_i^*, - \xi_i, \eta_i^*, - \eta_i ).
\end{equation}

 Now, one can decouple the interaction term by performing a Hubbard-
Stratonovitch transformation, which introduces a functional integral
over a matrix $Q$ which has the same symmetries as the dyadic product
$\psi_i \otimes \bar{\psi_i}$.
\begin{eqnarray}
\bar{C}_N(E, \omega) &=& \int d\psi d Q
\exp \left(\frac{i}{2} \sum_{i=1}^N
 \bar{\psi}_i (E + { \omega \over 2}\Lambda ) \psi_i \right)
\nonumber \\
&&\exp \left(-4c {\rm Tr} Q^2 + {\rm Tr} [(a_1 Q + a_2 \tau_3 Q \tau_3
) \sum_{i=1}^N
\psi_i \otimes \bar{\psi}_i ] \right),
\end{eqnarray}
 where $a_{1,2}= 1/2 ( \sqrt{1 +\alpha^2} - (-)^{1,2}
\sqrt{1-\alpha^2} )$.

Next, the integral over the vector $\psi$ can be performed, yielding
\begin{equation}
\int dQ \exp \left( - 4 c {\rm Tr} Q^2 \right)
\det \left( a_1 Q + a_2 \tau_3 Q \tau_3 + \frac{i}{2}(E+ { \omega \over 2}
 \Lambda\right)^{N/2}.
\end{equation}
 where we put now $c= \pi^2/4 N$, explicitly.

 Transforming $Q \rightarrow 2 \pi/N Q $, one obtains:
\begin{equation}\label{rmt}
\bar{C}_N(E, \omega) = \int dQ \exp \left[ - \frac{N}{4} {\rm Tr}
Q^2 + \frac{N}{2} {\rm Tr} \ln (\frac{N}{2 \pi} (a_1 Q + a_2 \tau_3 Q
\tau_3) + \frac{1}{2} i ( E + { \omega \over 2} \Lambda)) \right] .
\end{equation} Variation with respect to Q yields
${\rm Tr} \delta Q Q = {\rm Tr} (a_1 Q + a_2 \tau_3 Q \tau_3)^{-1} (
a_1 \delta Q + a_2 \tau_3 \delta Q \tau_3 ) $.  To lowest order in the
parameter $\alpha$ this saddle point condition becomes $Q^2 = 1$.
Expanding to lowest order in $\alpha$ and transforming $ Q \rightarrow
Q + \pi/N i ( E + { \omega \over 2} \Lambda )$ one obtains:
\begin{equation}\label{Evanish}
\bar{C}_N(E, \omega) = \int d Q \exp ( - N/4 {\rm Tr} Q^2 ) \det Q^{N/2}
\exp( \frac{\pi}{2} i {\rm Tr} [ (E + { \omega \over 2} \Lambda ) Q ] )
\exp ( \frac{N}{2} \alpha^2 {\rm Tr} [ Q \tau_3 Q \tau_3 ] ).\label{crossover}
\end{equation}
together with the symmetry restrictions of Q, $ Q = \bar{Q} $ where
$\bar{Q} = C Q^T C^T$ with $C = \left( \begin{array}{cc} -i\tau_2 & 0
\\ 0 & i \tau_2 \end{array}\right)$, with $\tau_2 = \left(
\begin{array} {cc}
0 & -i \\ i & 0 \end{array}
\right) $
and $ Q = Q^+ $ one obtains:
\begin{equation}
Q = \left( \begin{array}{cc} q \openone & A \\ A^+ & - q \openone
\end{array} \right).
\end{equation}
with $A= \left( \begin{array}{cc} a & b \\ b^* & -a^* \end{array} \right)$
where $ \mid a \mid^2 +
\mid b \mid^2 +  q^2 = 1$.
In the saddle point approximation $\bar{C}$ is independent of $E$,
 since $ Tr Q = 0$ for the above saddle point manifold.  Hence, the
 exponent in Eq. (\ref{Evanish}), is independent of $E$.  For $N
 \rightarrow \infty$ this saddle point approximation becomes exact.

 A matrix Q with the above symmetries can be represented as,
\begin{equation}
Q= U^{-1} Q_c^0 U,
\end{equation}
with
\begin{equation}
U = V_C U_D,
\end{equation}
where
\begin{equation}
U_D = V_D^{-1} T_D^0 V_D,
\end{equation}
where
\begin{equation}
 Q_c^0 = \left( \begin{array}{cc} \cos \theta_C & i \sin \theta_C \tau_2
\\  i \sin \theta_C \tau_2  & - \cos \theta_C \end{array} \right),
\end{equation}
and
\begin{equation}
T_D^0 = \left(\begin{array}{cc} \cos \theta_D/2 & i
\sin \theta_D/2 \\ i \sin \theta_D/2 & \cos \theta_D/2
\end{array} \right).
\end{equation}
and
\begin{equation}
V_{C,D} = \left(\begin{array}{cc} \exp ( i \phi_{C,D} \tau_3 ) & 0
 \\ 0 & \openone
\end{array} \right).
\end{equation}
and $\tau_i, i=1,2,3$ are the Pauli matrices.  Such a representation
 was first given by Altland, Iida and Efetov\cite{altland} to study
 the crossover between GOE and GUE within the supersymmetric nonlinear
 sigma model. Here, of course we need only the compact block of the
 representation given there.

 In order to perform the functional integral in this representation,
we have to find the corresponding integration measure.  To this end we
find $ Tr dQ^2 = ( d \theta_C d \theta_D d \phi_C d \phi_D ) A ( d
\theta_C d \theta_D d \phi_C d \phi_D )^T$ for this representation, and
making use of $ d Q = ( det A )^{1/2} d \theta_C d \theta_D d \phi_C d
\phi_D $, we obtain finally:
\begin{equation}
d Q = 16 \cos \theta_C^2 \mid \sin \theta_C \mid
\mid \sin \theta_D \mid d \theta_C d \theta_D d \phi_C d \phi_D.
\end{equation}

 As a result, we find for large N the normalized correlation function:
\begin{equation}\label{general}
C_N( E, \omega) = C_N( \omega) =  \int_0^1 d \lambda \lambda
\frac{\sin ( \pi \omega \lambda )}{\omega}
 \exp ( t^2 ( \lambda^2 -1 ))/C_0(t^2)
\end{equation}
where  the normalization constant is given by
\begin{equation}
C_0(t^2) = \frac{2}{ \pi} \int_0^1 d \lambda \lambda^2
\exp ( t^2 ( \lambda^2 -1)),
\end{equation}
 and $ t^2 = 4 N \alpha^2$.

  The simplest case is the pure GUE with $\alpha=1$.  Then, for large
N, the exponential factor in the integrand is finite only for $
\lambda =1$, and we obtain:
\begin{equation}
C_N(\omega)=
 \sin(\pi \omega)/ ( \pi \omega).
\end{equation}

 For $\alpha =0$, GOE, we find:
\begin{equation}\label{goe1}
C_N(\omega) = {3\over\pi} \int_{0}^{1} d \lambda \lambda
\frac{\sin ( \pi \omega \lambda)}{\omega}  = -  \frac{6}{ \pi^2 \omega}
\partial_{\omega} \frac{\sin(\pi \omega)}{\pi \omega}.
\end{equation}

 It is interesting to note that this coincides also with
\begin{equation} {3\over 2}
\int_{-1}^1 d \lambda ( 1 - \lambda^2 ) \exp ( i \omega \lambda)
\end{equation}
 as one obtains using Efetovs representation of the pure ensemble GOE.

For the general case $\alpha$ is arbitrary, and one can analytically
continue the integral (\ref{general}) and express it in terms of the
Error function with complex argument:
\begin{equation}
C_N(\omega) = \frac{\pi^{3/2} \frac{1}{t} \exp ( \frac{\pi^2
 \omega^2}{4 t^2} ) \Im [ {\rm erfc} ( \frac{\pi \omega}{2 t} + i t )]
 - 2 e^{t^2} \frac{\sin ( \pi \omega )}{ \omega} } {\pi^{3/2}
 \frac{1}{t} \Im [ {\rm erfc} ( i t )] - 2 e^{t^2} \pi }
\end{equation}
where ${\rm erfc} ( z )$ is the complementary Error function.
    

\section {Semiclassical Theory}

 The semiclassical quantization scheme which is to be used in the
sequel, is the scattering approach \cite {DoronUS}, (see also \cite
{Bogomolny}) in which one defines a semiclassical {\it unitary} $S(E)$
operator of dimension $\Lambda$ and the secular equation, defined to
be real for real energies, can be written as
\begin{equation}
 Z(E) = e^{ -i\Theta (E) /2 }\det (I-S(E)) \label {scatZ} 
\end{equation}
Where $\Theta (E)$ is the total phase shift:
\begin{equation}
 e^{ i\Theta (E) }= \det (-S(E)) .  
\end{equation}
It is assumed that the classical analogue of $S$ is an area preserving
map $\cal{M}$ acting on a Poincar\'e section with a phase space area
$A$.  In the semiclassical limit, $\Lambda$ is the integer part of $
{A\over 2\pi \hbar}$.  In the same limit, the total phase is related
to the smooth spectral counting function
\begin{equation}
\Theta (E) \approx 2\pi {\bar N(E)} \label {nbar}
\end {equation} 
The secular equation is real by construction. It can also be expressed
in terms of the eigenphases $\theta_l(E)$ of $S(E)$, or as the
characteristic polynomial of $S(E)$ evaluated at $z=1$:
\begin{equation}
 Z(E) = e^{ -i\Theta (E) /2 }\left [ \Pi _{l=1}^{\Lambda} (1-z
 e^{i\theta_l(E)})\right ]_{z=1} = e^{ -i\Theta (E) /2 } \left [
 \sum_{l=0}^{\Lambda} a_l(E) z^l \right ]_{z=1}
\label {scatZ1} 
\end{equation} 

The unitarity of $S(E)$ leads to the relations \begin{equation}
 e^{-i\Theta/2} a_l = e^{i\Theta/2} a_{\Lambda-l}^{*}
\label {scatZsymm} 
\end{equation}
 which can be used to rewrite (\ref {scatZ1}) as
\begin{equation}
 Z(E) = e^{ -i\Theta (E) /2 } \sum_{l=0}^{[\Lambda /2 ] } a_l(E) +{\rm
  c.c.}
\label {scatZ2} 
\end{equation}
 To calculate the correlation function, we remember that the $a_l$ are
the fully symmetric, homogeneous polynomials in the $e^{i\theta_m}$ of
order $l$. We can approximate their variation with energy by writing
\begin{equation} a_l(E+\epsilon/2) \approx a_l(E) e^{i l\tau
\epsilon/2 }
\end{equation}
 where $\tau$ is the average value of the partial delay times
$\tau_l={\partial \theta_l(E)\over \partial E}$. The distribution of
the $\tau _l$ is known to be narrowly centered about the mean value,
$\tau$ \cite {DoronUS}.  It follows from (\ref{nbar}) that
\begin{equation}
\tau = 2\pi{\bar d}/\Lambda 
\end {equation}
The autocorrelation function reads now
\begin{eqnarray}
{\bar C(\epsilon)}& = &{1\over \Delta E} \int_{E_0-\Delta
E/2}^{E_0+\Delta E/2} {\rm d}E <Z(E+\epsilon/2)Z(E-\epsilon/2)> \\
\nonumber &\approx & \sum _{l=0}^\Lambda |<|a_l|^2>_{\Delta E} e^{i
\tau \epsilon(\Lambda /2-l) }
\label {scatcorrl}
\end{eqnarray}
Where we made use of the fact that the interval $\Delta E$ is large on
the quantum scale so that the phases $\theta_l(E)$ make many
revolutions when $E$ traverses the interval $\Delta E$.  We may now
denote the scaled energy by $\omega = \epsilon \bar d$, and use the
relation $\tau \epsilon = {2\pi\over \Lambda} \epsilon {\bar d} =
{2\pi\over \Lambda}\omega $. The last equation shows that the
autocorrelation function $\bar C (\epsilon)$ can be also interpreted
as the autocorrelation of the secular equation of the ensemble of
$S(E)$ matrices averaged over the energy interval $\Delta E$. Hence,
in comparing the results of the semiclassical theory with the
predictions of RMT, we shall use the RMT expression for the {\it
circular} ensembles (\ref {corrcen}), even though our starting point
was the spectral determinants of the Hamiltonian.
 
To introduce the semiclassical theory for the autocorrelation
function, it is useful to recall some exact relations which enable us
to express the coefficients $a_l$ in terms of ${\rm tr} S^n$. This is
done by iterating the Newton identities \begin{equation} a_l =-{1
\over l} \left ({\rm tr} S^{l}+ \sum _{k=1}^{l-1} a_k {\rm tr} S^{l-k}
\right ).
\label {newton}
 \end {equation} An explicit solution of the Newton identities is
 given by the {\it Plemelj-Smithies formula}
\cite {Plemelj}
\begin {eqnarray} 
 a_l = {(-1)^l\over l!} 
\begin {array} {c}\left |
 \ \begin {array} {lllllll} s_1 \ \ & s_2 \ \ & s_3\ \ & \cdot \ \ \ \
 & \cdot\ \ \ \ &\cdot \ \ & s_l \\ 1 & s_1 & s_2& \cdot & \cdot
 &\cdot & s_{l-1} \\ 0 & 2 & s_1& s_2 &\cdot & \cdot & s_{l-2} \\ 0 &
 0 & 3 & s_1 & \cdot &\cdot & s_{l-3} \\
\cdot &\cdot &\cdot &\cdot &\cdot &\cdot &\cdot       \\
\cdot &\cdot &\cdot &\cdot &\cdot &\cdot &\cdot      \\
0    &\cdot &\cdot &\cdot &0 &l-1 & s_1 
\end {array}
\right | \label{fredholm}
\end {array}  
\end{eqnarray}
Where we use the notation 
\begin{equation}
 s_l = {\rm tr} S^{l}
\end{equation}
The result (\ref {fredholm}) can be easily proved by expanding the
determinant with respect to the last column.  Writing down the
explicit expression of the determinant, one gets
\begin {equation}
a_l = -{1 \over l}\left (s_l + \sum _{\vec l} {(-1)^{n}\over \prod_{i=1}^n l_i}
s_{l-l_1} s_{l_1-l_2}\cdots s_{l_{n-1}-l_n}s_{l_n} \right )
\label {fredholm2}
\end{equation}
where the summation is over all vectors $\vec l$ with integer entries
such that $l>l_1>l_2> \cdots >l_n \geq 1$. The last equation can also
be directly derived from (\ref {newton}) by successive iterations.
    
 The semiclassical approximation is introduced at this point: The $S$
operator is considered as the quantum analogue of the classical
evolution operator of the relevant Poincar\'e map $\cal {M}$. Hence
${\rm tr} S^n$ is expressed semiclassically in terms of periodic
orbits which traverse the Poincar\'e section $n$ times.
\begin{equation}
s_n = {\rm tr} S^n \approx \sum_{p} {g_p n_p e^{ir (S_p/\hbar-\nu_p
              {\pi\over 2})} \over |\det (I-M_p^r)|^{{1\over 2}} }
\label{trsemicl}
\end{equation} 
The summation is extended over all primitive periodic orbits $p$ of
the Poincar\'e map, with periods $n_p$ which are divisors of $n$, so
that $ n=n_p r$. $g_p$ stands for the number of distinct symmetry
conjugate orbits, $M_p$ is the monodromy matrix, $S_p$ and $\nu _p$
are the action and the Maslov index, respectively.

To get the semiclassical expression for the coefficients of the
autocorrelation function, we have to substitute the semiclassical
expression (\ref {trsemicl}) in (\ref {fredholm2}). When this is done,
one gets the standard {\it composite orbits} expansion of the spectral
determinant \cite {DoronUS}\cite {KeatingBerry}.  Taking the absolute
square and averaging over the energy interval, we make use of
the diagonal approximation,
\begin{equation}
 < s_n s^* _m >_{\Delta E} \approx \delta _{nm} \sum_{p} { g_p^2 n_p^2
\over |\det(I-M_p^r)|}
\approx  \delta _{nm} \sum_{p} { g^2 n^2 \over |\det(I-M_p)|} 
\label {diagonal}
\end{equation}   
 Which is valid because of the rapid oscillating phases in (\ref
{trsemicl}). The right expression in (\ref {diagonal}) makes use of
the observation that for large $n$ the primitive orbits dominate the
periodic orbit sum. In the same spirit, $g_p$ is replaced by its mean
value $g$. In the following we shall use the notation
\begin{equation}
 <|s_n|^2> = 
 \sum_{p} { g^2 n^2 \over |\det(I-M_p)|} 
\label {diagonalnota}
\end{equation}     
The first step in implementing the diagonal approximation for $<
|a_l|^2>_{\Delta E}$ is to isolate terms, which upon averaging, do not
yield vanishing contributions because of unmatched phase factors. Thus
\begin{eqnarray}
\label {asqav} 
 < |a_l|^2>_{\Delta E} &=& {1\over l^2}\left[ <s_l s^*_l>_{\Delta E} +
< \sum_{l_1}\sum_{m_1} {s_{l-l_1}s_{l_1}s^*_{l-m_1}s^*_{m_1} \over l_1
m_1}>_{\Delta E}
\right. \\ 
&+&
\left. < \sum_{l_1,l_2} \sum_{m_1,m_2} 
{s_{l-l_1}s_{l_1-l_2}s_{l_2}s^*_{l-m_1}s^*_{m_1-m_2}s^*_{m_2} \over
 l_1 l_2 m_1 m_2 }>_{\Delta E} + \cdots \right ] \nonumber
\end {eqnarray}
Where the summations in the $n'$th term are restricted to the domains
$l>l_1>l_2> \cdots >l_n \geq 1$ and $l>m_1>m_2> \cdots >m_n \geq
1$. The diagonal approximation for the first term in (\ref{asqav}) is
our building block (\ref {diagonalnota}):
\begin{equation}
<s_l s^*_l>_{\Delta E} = <|s_l|^2>
\end {equation}
 In the next  $(n=1)$ term we can pair the indices in two ways
which will survive the energy averaging:  either $m_1=l_1$ or $m_1=l-l_1$.
\begin{equation}
< \sum_{l_1}\sum_{m_1} {s_{l-l_1}s_{l_1}s^*_{l-m_1}s^*_{m_1} \over l_1
m_1}>_{\Delta E} = l \sum_{l_1} {1\over l_1} {<|s_{l-l_1}|^2> \over
l-l_1} {<|s_{l_1}|^2> \over l_1}
\end{equation}
The implementation of the diagonal approximation to the next terms
enforces us to choose the indices $m_k$ so that the set of differences
$\{ m_k -m_{k+1}\}$ is a permutation of the set of differences
$\{l_k-l_{k+1}\}$. Hence, in the general case, we have $(n+1)!$ non
vanishing contributions.  To understand how they sum up, consider the
$(n=2)$ term for which there are already 6 ways to chose the
indices. The six resulting contributions and their sum are
\begin{eqnarray}
\lefteqn{< \sum_{l_1,l_2} \sum_{m_1,m_2} 
{s_{l-l_1}s_{l_1-l_2}s_{l_2}s^*_{l-m_1}s^*_{m_1-m_2}s^*_{m_2} \over
 l_1 l_2 m_1 m_2 }>_{\Delta E}= \sum_{l_1,l_2}
 {<|s_{l-l_1}|^2><|s_{l_1-l_2}|^2> <|s_{l_2}|^2> \over l_1l_2} \times
 } \nonumber \\ & &\left({1\over l_1l_2}+{1\over l_1(l_1-l_2)}+{1\over
 l_2(l-l_1+l_2)} + {1\over (l-l_2)(l_1-l_2)} + {1\over (l-
 l_1)(l-l_1+l_2)} +{1\over (l-l_1)(l-l_2)} \right) \nonumber \\
 \nonumber \\ & & = l\sum_{l_1,l_2} {1 \over l_1 l_2} {<|s_{l-l_1}|^2>
 \over l-l_1} {<|s_{l_1-l_2}|^2> \over l_1-l_2} {<|s_{l_2}|^2>\over
 l_2} \ \ .
\end{eqnarray}
 The $n=3$ with 24 contributions can be also worked out explicitly,
resulting in a simple expression. This can be generalized to any value
of $n$, and when it is substituted in (\ref{asqav}) we get,
\begin{equation}
< |a_l|^2>_{\Delta E} \approx {1 \over l} \left[ {<|s_l|^2> \over l} +
\sum _{n=1}^{l-1} \sum_{l_1\cdots l_n }{1\over \prod_j l_j}  
{ <|s_{l-l_1}|^2> \over l-l_1} {<|s_{l_1-l_2}|^2> \over l_1-l_2}
\cdots {<|s_{l_n}|^2> \over l_n}
\right] \ \ ,
\label {afinal}
 \end{equation} where the summation is carried with the restriction
$l>l_1>\cdots \l_n \geq 1$.

 At this point it is instructive to consider the Fredholm determinant
for the {\it classical} evolution (Frobenius Peron) operator
\begin {equation}
U({\bf x,x'}) = \delta ({\bf x'}-{\cal{M}}({\bf x})),
\end{equation}
where ${\bf x}$ is a point on the classical Poincar\'e section. A
straight forward integration gives
\begin{equation}
u_l = {\rm tr}U^l = \sum_{p_l} {l_p g_p \over |\det(I-M_p^r)|} \approx 
\sum_{p}{l g \over |\det(I-M_p)|}  = {1\over gl}<|s_l|^2>
\end{equation}
 The uniform coverage of phase space by the chaotic trajectories
implies that $u_l \approx 1$ for sufficiently large $l$. In other
words,
\begin{equation}
<|s_l|^2> \approx gl  \ \  \ \ \ {\rm for } \ \ l>1
\end{equation}
which is the well known Hannay and Ozorio de Almeida sum rule \cite
 {hannay}.  Using this estimate for all $ <|s_l|^2> $ in (\ref
 {afinal}) we find
\begin{eqnarray}
\label {almostthere}
 < |a_l|^2>_{\Delta E} \approx {g\over l}\left (1+ \sum _{n=1}^{l-1}
 g^n \sum_{l>l_1\cdots l_n\geq 1 }{1\over \prod_j l_j}\right )
 ={g\over l} \left(1+\sum _{n=1}^{l-1} g^nI_n(l) \right ) \ \ ,
\end{eqnarray}
where
\begin{equation}
I_n(l) = \sum_{l_1=n}^{l-1}{1\over l_1}\sum_{l_{2}=n-1}^{l_1}{1\over
l_{2}}
\cdots \sum_{l_{n-1}=2}^{l_{n-2}}{1\over 
l_{n-1}}\sum_{l_n=1}^{l_{n-1}}{1\over l_n}
\end{equation}
The lower and upper summation indices follow from the restriction
 $l>l_1>\cdots \l_n \geq 1$. We now observe that the functions
 $I_n(l)$ satisfy the recursion relations
\begin{equation}
I_n(l)=-{1\over l}I_{n-1}(l) +\sum_{r=n}^{l} {1\over r}I_{n-1}(r)
\label {recur1}
\end{equation}
which is equivalent to 
\begin{equation}
I_n(l)-I_n(l-1)={1\over {l-1}}I_{n-1}(l-1) 
\label {recur2}
\end{equation}  
and it is subject to the condition $I_l(l)=I_l(l-1) =0$. Denote 
\begin{equation}
f_l(g) = 1+ \sum _{n=1}^{l-1} g^nI_n(l) 
\end{equation}
Multiplying (\ref{recur2}) by $g^n$ and summing over $n$
one gets
\begin{equation}
f_l(g)={l-1+g\over {l-1}}f_{l-1}(g) \ \  \ {\rm with}  \ \  \ f_1(g) =1 \ \ .
\label{recur3}
\end{equation}
Hence, for any integer $g$  
\begin{equation}
f_l(g)= {(l-1+g)!\over (l-1)!g!} = \left( \begin{array}{c}l-1 +g\\ g
\end{array} \right) \\ .
\end{equation}
For the cases of interest here, $f_l(g=1) = l$, and, $f_l(g=2) =
l(l+1)/2 $.  These expressions can be substituted in
(\ref{almostthere}) to give
\begin{equation}
 < |a_l|^2>_{\Delta E}(g) = {g\over l} f_l(g) = \left\{
 \begin{array}{c} 1 \ \ \ \ \ \ \ {\rm for} \ \ g=1 \\ 1+l \ \ \ {\rm
 for} \ \ g=2 \end{array} \right.
\label{mainresult}
\end{equation}
 
This is the central result of the present chapter. It should be
augmented by the identity
\begin {equation} \label {symmmm}
 < |a_l|^2>_{\Delta E}= < |a_{\Lambda-l}|^2>_{\Delta E}
\end{equation}
which is due to unitarity, and then compared with the results of RMT
  (\ref {haake}).

Systems {\it without} time reversal symmetry (TRS) have $g=1$, so that
the semiclassical result (\ref {mainresult}) coincides with the
prediction of RMT for the CUE.

 In chaotic systems {\it with} TRS, $g=2$ and
\begin{eqnarray}
\label {sclgoe}
< |a_l|^2>_{\Delta E}\ \ \approx \ \ \left\{ \begin {array}{l} 1 + l \
 \ \ \ \ \ \ \ \ \ {\rm for }\ \ 1<l \leq \Lambda /2 \\ 1+\Lambda -l \
 \ \ \ {\rm for}\ \ \Lambda>l \geq \Lambda/2 \end{array} \right. .
\end{eqnarray}
 
 This expression does not reproduce the RMT result for the COE case
(\cite {haake})
\begin{equation}
< |a_l|^2>_{COE} \ \ = \ \ 1+l{\Lambda \over \Lambda +1} -l^2{1 \over
\Lambda +1}
\end{equation}
 However, for large $\Lambda$, where the semiclassical approximation
is justified, the semiclassical result reproduces the exact expression
in a domain of $l$ values of size $\sqrt{\Lambda}$ in the vicinity of
the end points of the $l$ interval, $l=0$ and $l=\Lambda$.  The
deterioration of the quality of the agreement between the
semiclassical and the RMT expressions when TRS is imposed is typical,
and it is an enigma in the field of quantum chaos.  This is a typical
behavior of the semiclassical approximation and it was discussed first
by Berry \cite {BerryA400}.

We shall use now the results of the previous sections to investigate
the connection between the spectral autocorrelation function and the
Ruelle $\zeta$ function for the classical mapping $\cal M$. The Ruelle
$\zeta$ is constructed from the Fredholm determinant of the classical
evolution (Frobenius Peron) operator by
\begin{equation}
\zeta (s) = \left(\det (I-e^{-s} U) \right)^{-1} = 
\sum_{l=0}^{\infty} A_l^{\rm cl}e^{-sl} 
\end {equation}
Using the same methods as above, one can easily derive the explicit
expression for the coefficients,
\begin {equation}
A_l^{\rm cl} = {1 \over l}\left (u_l + \sum _{\vec l} {1\over
\prod_{i=1}^n l_i} u_{l-l_1} u_{l_1-l_2}\cdots u_{l_{n-1}-l_n}u_{l_n}
\right )
\label {fredholmclass}
\end{equation}
and compare it to the coefficients of the semiclassical correlation
function (\ref{afinal}) in which the relation $s_l \approx g l u_l $
is used,

 \begin{equation} \label {bbb} < |a_l|^2>_{\Delta E} \ \ \approx \ \
{g\over l}\left( u_l + \sum _{\vec l} g^n{1\over \prod_{j=1}^n l_j}
u_{l-l_1} u_{l_1-l_2} \cdots u_{l_n-l_{n-1}} u_{l_n}\right).
\end{equation} For systems without TRS $g=1$, and one gets
\begin{equation}
< |a_l|^2>_{\Delta E} \approx A_l^{\rm cl}
\end{equation}
 This close relationship between the {\it quantum} autocorrelation
function and the {\it classical} Ruelle $\zeta$ function is another
manifestation of the observation of
\cite {AAA} and \cite {Bogolkeating} on the r\^ole played by the 
Ruelle $\zeta$ function in the semiclassical theory of spectral
correlations. However, one has to be careful and modify this statement
by noting that only a {\it finite} number of coefficients are used in
the spectral autocorrelation function. As a matter of fact, only the
first $\Lambda/2$ are used.  Those with $ \Lambda/2 \leq l \leq
\Lambda$ must be defined by (\ref {symmmm}) due to unitarity, and the
rest are set to zero. This is where the (topological) Heisenberg time
enters the semiclassical theory. It should be emphasized that the
truncation and the symmetry are essential elements, without which the
semiclassical theory yields wrong results.  Therefore, the
identification of $C(w)$ as $\zeta (i\pi w)$ is not allowed.

 System with TRS have $g=2$. In this case one can identify the
coefficients $< |a_l|^2>_{\Delta E}$ with the coefficients of the
function
\begin{equation}
\zeta_{g}(s) = \left(\det (I-g e^{-s} U) \right)^{-1} = 
\sum_{l=0}^{\infty} A_l^{\rm cl} (g) e^{-sl}
\end{equation}  
For large $s$ one can use the approximate relation $\zeta_g(s) \approx
(\zeta (s))^g $. The fact that $(\zeta (s))^2$ is the classical
function which appears in the semiclassical approximation for spectral
statistics in systems with TRS was previously discussed in \cite
{Bogolkeating}. The remarks made above concerning the relation between
the autocorrelation function $C(\omega)$ and the classical function
$\zeta_g(s)$ hold in the present case as well.

 The explicit results obtained above are valid under the condition
that the classical orbits uniformly cover phase space, and uniformity
is achieved within a short time. Only for such systems $u_n \approx
1$. There are, however, other chaotic systems for which the coverage
of phase space is diffusive. Then, $u_n$ can be identified with the
{\it classical return probability}, which for diffusion in $d$
dimensions is proportional to $n^{-d/2}$. One can work out the
spectral autocorrelation function for these systems as well. This
subject will be pursued elsewhere.

\section{Numerical Results and Illustrations}

In the previous sections we discussed the average autocorrelation
 functions of spectral determinants for various random matrix
 ensembles, and the corresponding semiclassical expressions. Before
 being able to compare the RMT results with spectral data of quantum
 chaotic systems, we had to clarify a few practical points.  The
 results for the Gaussian ensembles are obtained in the large $N$
 limit, while all the numerical work can be carried out on spectral
 intervals of {\it finite} length.  Since the differences between the
 correlation functions for the different ensembles are not too large,
 it was important to develop tools to check to what extent the finite
 $N$ calculations approach the $N\to\infty$ limit.  For the GOE case,
 the analytic theory becomes rather cumbersome.  The analytic results
 of \cite {haake} for the Circular ensembles cannot be used directly
 for this purpose, because the Circular and the Gaussian ensembles are
 expected to coincide only in the $N\to\infty$ limit.  To check this
 point, we developed an efficient sampling method based on the
 Metropolis algorithm \cite{Metropolis}, with which we could calculate
 the expectation values of any property for any of the matrix
 ensembles both accurately and efficiently.  This was an important
 tool in our work and we shall describe it shortly in the first part
 of this chapter.

 Spectral determinants of unbounded Hermitian operators do not
converge unless properly regularized. In practical applications, one
has to make a choice of the regularization method. In the second part
of this chapter we discuss two alternative approaches and show how
they are applied to the analysis of long spectral intervals of Sinai
billiards in two and in three dimensions.

\subsection{ The Metropolis Algorithm for RMT }

 The Metropolis Algorithm (MA) generates a {\it finite} ensemble of
 spectral sequences which are statistically independent, and which are
 distributed according to a prescribed probability distribution. This
 finite ensemble is used to perform averages which are guaranteed to
 converge to the exact ensemble averages, when the number of members
 in the ensemble increases.

 The MA for the circular ensembles is defined by the $N$ point
spectral probability distribution
\begin{equation}
p( \vec \theta ) = {\cal N} \prod_{1\le k<l \le N} | e^{ i \theta_k}-
 e^{i \theta_l} | ^\beta \label{probdis}
\end{equation}
were $\vec \theta =\{\theta_1, ..., \theta_N\} \ \ \ , \ \ \ 0\leq
 \theta_l \leq 2\pi$, and ${\cal N}$ is a normalization constant.  The
 MA proceeds as follows.  Starting from an initial vector $\vec
 \theta_n$ a new vector $\vec \theta_{n+1}= \vec \theta_n + \vec
 \Delta \theta_n$ is determined with the components of $\vec \Delta
 \theta_n$ randomly chosen from $[-\Delta \theta_{Max}:\Delta
 \theta_{Max}]$.
\begin{figure}[htb]
\begin{center}
\mbox{\psfig{figure=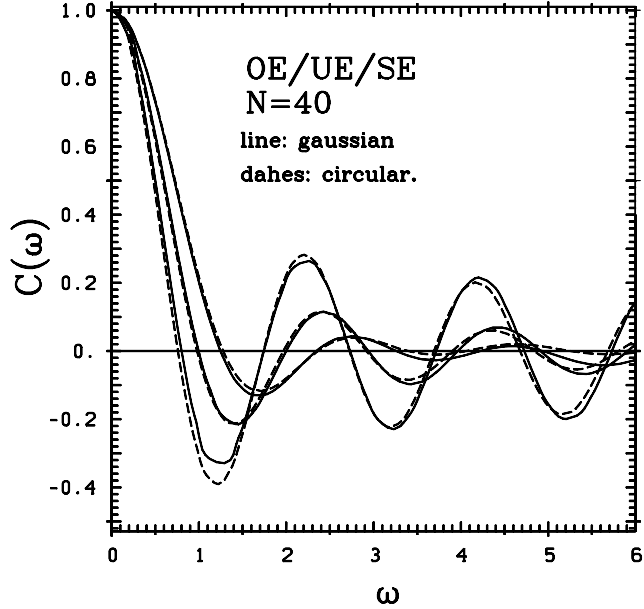,height=8cm}}
\end{center}
\caption{\sl $C(\omega)$ for Gaussian (line) and circular (dashes) ensemble.
  Results are shown for the orthogonal (most strongly damped) the
  unitary (middle) and symplectic ensemble (least damped). }
\label{gauscirc}
\end{figure}

Denote
\begin{equation}
r=p( \vec \theta_{n+1} )/p( \vec \theta_{n} ) \quad .
\end{equation}
A step is accepted if $r>1$ or $\eta<r$ where $\eta\in[0,1]$ is a new
random number at each step.  As the algorithm depends only on the
ratio $r$, the normalization constant is irrelevant.  The choice of
$\Delta \theta_{Max}$ is crucial for near optimal performance of the
algorithm and we use $\Delta \theta_{Max}=0.02$ for $N=40$ and $\Delta
\theta_{Max}=0.007$ for $N=80$. For these parameters, approximately
fifty percent of the Metropolis steps are accepted.  The initial
vector $\theta_0$ is chosen at random and the transients due to the
initial choice are erased after a few iterations since the "random
walker" $\vec \theta_n$ advances quickly towards the most probable
domain.  Out of all the vectors $\vec \theta_{n}$ only a small
fraction is accepted to the ensemble in order to obtain an
uncorrelated sample. Each run generates a set of 2000 properly
distributed vectors which are then used for further analysis.  We
checked the convergence with respect to the ensemble size, and found
that a set of 2000 spectra is sufficiently large to obtain reliable
results.

The same algorithm can be used to sample the probability measure of
 the Gaussian ensembles
\begin{equation}
p( \vec x )= {\cal N}e^{-{\beta \over 2} \sum_{i=1}^N x_i^2 }
\prod_{1\le i<j \le N}
| x_i- x_j |^\beta \label{progaus}
\end{equation}
 One has to exercise some  care  to avoid the choice of
initial conditions which have vanishingly small probability due to the
 exponential term
in (\ref {progaus}).

 As the first application, we studied the difference between the
 autocorrelation functions for the circular ensembles $(\beta =
 1,2,4)$ and their corresponding Gaussian ensembles for a {\it finite}
 dimension $N=40$.  We used the definitions (\ref {corr1}) and (\ref
 {corr}) for the autocorrelation functions.  For the circular
 ensemble, the domain of integration is $[0,2\pi]$ and for the
 Gaussian ensembles we limited the integration domain to the interval
 $ |E| \leq {\sqrt N}$ to avoid deformations whose origin is the
 nonuniform semicircular level density.  Fig.~\ref{gauscirc} shows
 small yet systematic differences between the Gaussian and Circular
 ensembles which are most pronounced for the Symplectic and the
 Orthogonal ensembles.  The differences between the Unitary ensembles
 is the smallest. The overall agreement is rather good, as expected
 when the limit $N \to \infty$ is approached.

\begin{figure}[htb]
\begin{center}
\mbox{\psfig{figure=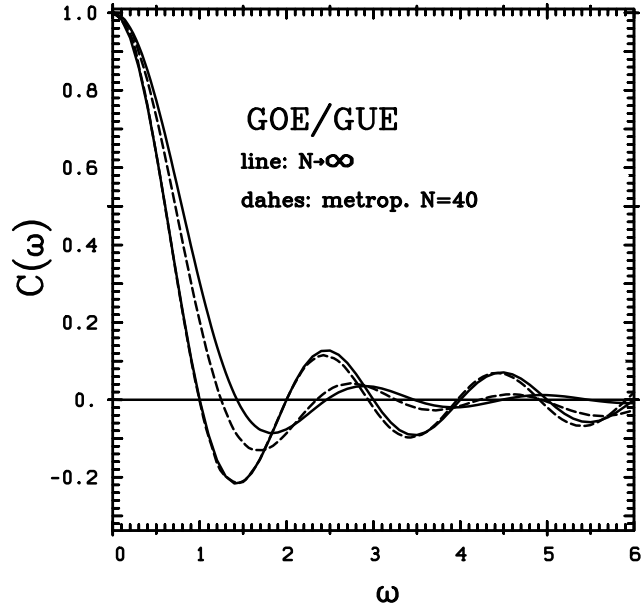,height=8cm}}
\end{center}
\caption{\sl Comparison of $N \to \infty$ results (line) and the  $N=40$
  (Metropolis) data for the GOE and GUE }
\label{exgaus}
\end{figure}

 The same trend persists also when one compares the $N=40$ cases for
the Gaussian ensembles with the expressions derived in section
~\ref{theory} for $N \to \infty$. This is shown in
Fig.~\ref{exgaus}. We see again that the $N=40$ data is fairly close
to the infinite limit for the GUE case. In the GOE case the $N=40$
data is significantly different from the $N \to \infty$ limit.

 We used two independent ways to treat a given large spectrum. In the
first method the large spectrum with $N_{total}$ eigenvalues is
"chopped" to a number of smaller subintervals. Given a large spectrum,
one can chop it in various ways, and one has to find a compromise
which provides a sufficiently large ensemble of not too short
intervals. We worked with $\approx N_{total}/40$ subintervals of
$N=40$ eigenvalues . Each subinterval, is unfolded to a mean spacing
of 1 and is shifted to be centered around $E=0$. This procedure puts
the subintervals on an equal footing.  For the calculation of the
correlation function, the integration is limited to $|E|<\sqrt{N}$.
The ensemble averaged autocorrelation $\bar C(\omega)$ is normalized
by $\bar C( \omega = 0)$.  The calculation of the autocorrelation
function is similar to the way by which we used the MA samples which
were constructed according to the Gaussian measures.
An alternative definition would have been to consider the
 autocorrelation function of the spectral determinant built from the
 entire large spectrum. The definition ~(\ref{corr1}) will be harder
 to apply because both the numerator and the denominator diverge with
 either $N=N_{total}$ or $E$. This difficulty is related to the well
 known problem of regularization of spectral determinants, and the
 regularization scheme we have chosen is the following. We shift the
 origin of the energy axis to the center of gravity of the large
 spectrum, and consider the secular function
\begin {equation}
Z_R(E) = \prod_{i=-N}^{N-1} \left(1-{E \over E_i}\right)
\end {equation}
This function is defined also in the limit of $N \to \infty$ as long
 as the symmetric construction is maintained in the limiting
 procedure. However, for finite $N$ the function $Z_R(E)$ diverge as
 $E^{2N+1}$ for sufficiently large $E$. This can be cured by a
 regularization factor whose origin can be illustrated by considering
 the spectral determinant for an equidistant spectrum:

\begin{equation}
p_R(E)=\prod_{i=-N}^{N-1} \left(1-{E\over i+{1\over2}}\right) \ \ =
\ \  { \Gamma(N+{1\over2}-E) \Gamma(-N+{1\over2})
\over   \Gamma(-N+{1\over2}-E) \Gamma(N+{1\over2}) }
\qquad . \label{harmpol}
\end{equation}

Using the asymptotic expressions for the $\Gamma $ function one gets
\begin{equation}
p_R(E)\approx \cos(\pi E) e^{ E^2 \over N} \label{harmasym}
\qquad .
\end{equation}
which is valid for $ |E| <E_{max}(N) $ where $\sqrt{N} < E_{max}(N) <<
N$.  This suggests the definition of a regularized corrected spectral
determinant
\begin{equation} \label{regcorpol}
Z_{\tilde R}(E)=e^{ - {E^2 \over N}}\prod_{i=1}^N (1-{E\over E_i})
\end{equation}
Where the $E_i$ are the unfolded, symmetrically centered spectrum.

The autocorrelation function can now be calculated using
\begin{equation} \label{regsin}
\bar C_{\tilde R}(\omega)= \int_{-{\tilde N}}^{\tilde N }
Z_{\tilde R}(E-{\omega\over2}) Z_{\tilde R}(E+{\omega\over2}) dE
\qquad.
\end{equation}
Note that all eigenvalues are used in one run.  Here $\tilde N$ is
proportional to $N$, but smaller than ${N \over 2}$ as even the
regularized spectral determinant is much larger than 1 at the edges of
the spectrum. We arbitrarily used $\tilde N = {N \over 6}$ which
yields good results.

\begin{figure}[htb]
\begin{center}
\mbox{\psfig{figure=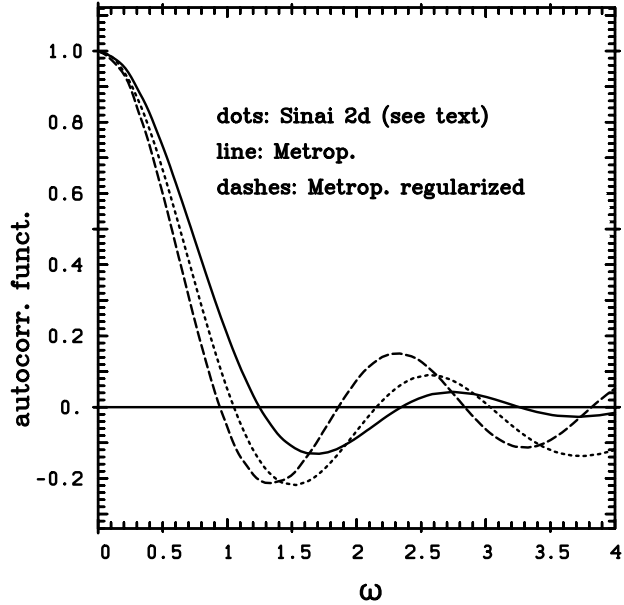,height=8cm}}
\end{center}
\caption{\sl line: autocorrelation function  $C(\omega)$ for GOE
  using Metropolis data for $N=40$; dashes: like previous but
  regularized $C_R(\omega)$; dots: regularized and corrected
  $C_{\tilde R}(\omega)$ for the 2d Sinai billiard.}
\label{regpol}
\end{figure}

As long as $N$ is finite the correlation function based on the
regularized polynomial can be expressed as
\begin{equation} \label{regcor}
C_R(\omega)=<\bar C_R(\omega)>=<{ \bar C(\omega) \over \bar C(0)} > \qquad.
\end{equation}
Thus, the difference between the two definitions of the correlation
 function comes from the different order of the operations of
 averaging and normalizing.  In Fig.~\ref{regpol}, we compare the
 correlation functions obtained by the two averaging methods. The
 dashed and the solid lines are MA runs for $N=40$ with GOE
 statistics, using the original (\ref {corr}) and regularized (\ref
 {regcor}) definitions, respectively.  The regularized correlation
 function is less damped, and it resembles the non-regularized
 function for GUE. (The RMT expression for the correlation function
 (\ref {regcor}) for the GUE was considered in \cite {AS})

 We have emphasized several times in the discussion above that the
expectation value of the autocorrelation function depends on all the
$n$-point spectral statistics, and therefore does not, in principle,
contain any new information.  However, since a typical spectral
analysis is usually carried out in terms of only a few statistical
measures, it is worth while to check to what extent the information in
the autocorrelation function overlaps with the most common statistics
- the nearest neighbour distribution $P(s)$.
\begin{figure}[htb]
\begin{center}
\mbox{\psfig{figure=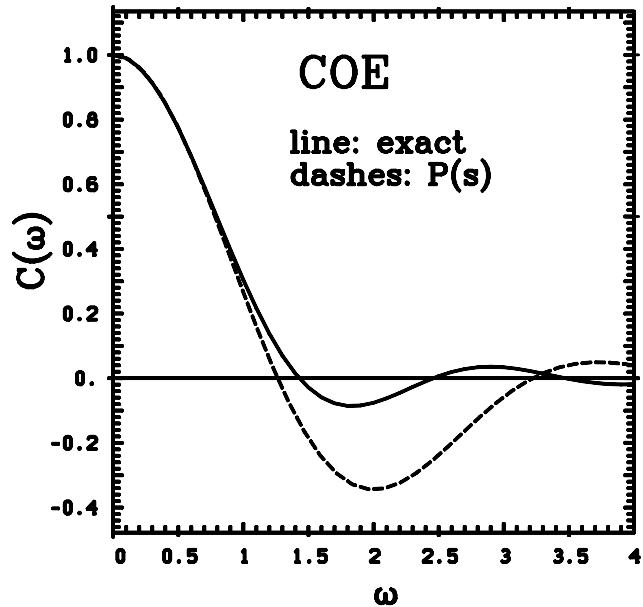,height=8cm}}
\end{center}
\caption{\sl  Comparison of the $N \to \infty$ results (line) and the
  results for spectra which only have the correct nearest neighbour
  level spacing distribution but higher order correlations are
  neglected.  (dashes; N=40). All results are for COE.}
\label{lespspec1}
\end{figure}
 For this purpose we used the rejection method to generate a random
 ensemble of spectral sequences, with nearest neighbour spacings which
 are distributed as $P_{COE}(s)$, disregarding all other correlations
 which are implied by the exact COE measure (\ref {progaus}).  In
 Fig. ~\ref {lespspec1} we compare the expectation value of the
 autocorrelation function for this ensemble, with the proper COE
 result. The difference between the two functions shows already at the
 positions of the first zero crossings, and the depths of the first
 minima. These differences are rather large, and establish the
 sensitivity of the autocorrelation function to correlations between
 levels which are further apart than the nearest neighbours.

\subsection{Application to Spectra of Physical Systems}

At this point we turn to the analysis of physical spectra, to check
the extent of applicability of the semiclassical and the RMT results.

\begin{figure}[htb]
\begin{center}
\mbox{\psfig{figure=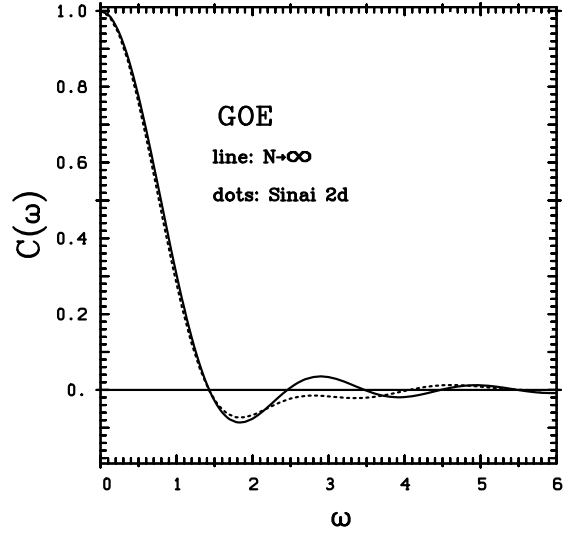,height=7cm}}
\end{center}
\caption{\sl  Autocorrelation function for the 2 dimensional 
 Sinai billiard. The spectrum is chopped into pieces with N=40. }
\label{sinai2}
\end{figure}

\begin{figure}[htb]
\begin{center}
\mbox{\psfig{figure=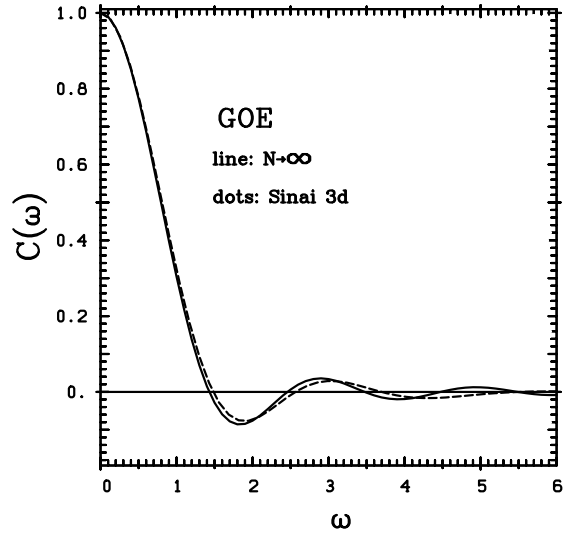,height=7cm}}
\end{center}
\caption{\sl Same as previous figure but for 3 dimensions.}
\label{sinai3}
\end{figure}

We used the spectra of Sinai billiards in two \cite {schanz} and in
 three dimensions \cite {primack}, for which we have the lowest 2591
 and 6698 eigenvalues, respectively. We analyzed the (unfolded)
 spectra using the two methods mentioned previously. In the first, the
 spectra were chopped into subintervals of length $N=40$, and were
 centered around $E\approx0$. We thus obtained numerical ensembles
 consisting of $64$ and $167$ intervals respectively.  The average
 autocorrelation functions are compared to the corresponding
 Metropolis data in Fig.~\ref{sinai2} and Fig.~\ref{sinai3}.  As the
 number of eigenvalues is quite small for the 2 dimensional Sinai
 billiard, only the first minimum is reproduced correctly. However,
 the situation is improved when the larger spectrum for the 3
 dimensional billiard is analyzed.

  To use the entire spectral data in one run, we defined the
correlation function as in (\ref{regsin}) and applied it to the
spectrum of the 2 dimensional Sinai billiard, using all eigenvalues in
one run.  The result is given in Fig.~\ref{regpol} (dots).  The
numerical values agree better with the MA runs for the regularized
version.  Both show nearly the same damping although the zeros are
slightly shifted.

The conclusion we may draw from the comparison of the two methods, is
that the regularized version has a built in self-averaging mechanism.
The energy integral represents an averaging over independent parts of
the spectrum. At a certain value $E$ only $E_i$ which are close to $E$
significantly contribute to the product (\ref{regcorpol}).

\section{Conclusion}

 In this work we tried to emphasize the theoretical and practical
advantages of the use of the autocorrelation function of spectral
determinants, both as a practical statistical tool and as an
alternative route to study the relationship between quantum chaos and
RMT. To conclude, we would like to make a few further comments.

 One of the most gratifying aspects of the present work was the fact
that the correlations of the Riemann $\zeta$ on the critical line have
much in common with the present work, and could be studied using
rigorous mathematical tools \cite {Keatingzeta}. This provides much
more confidence in the ad hoc assumptions which were made in the
development of the semiclassical theory.

 Our work on the RMT expressions for the autocorrelation functions
complements the previous results on the circular ensembles \cite
{haake}, and the the known results for GUE \cite {AS}. It also
provides the expressions for the GOE case, and for the ensembles which
interpolate between GOE and GUE.

 The semiclassical derivation of the autocorrelation function is new,
and enables a close scrutiny of relevance of the classical $\zeta$
function to spectral statistics in the semiclassical limit. It is
clear that both functions are derived form the same building blocks -
as a matter of fact, their Fourier coefficients coincide up to the
topological time which corresponds to half the Heisenberg time. This
is in agreement with the recent observations of \cite
{Bogolkeating}. The semiclassical methods which were developed here
can also be used for quantum chaotic problems for which the spectral
statistics differ from the one expected for the standard Dyson
ensembles. Such problems are commonly met in applications to
mesoscopic systems.
 
\section{Acknowledgments}
This work was supported by the Minerva Center for Nonlinear Physics
 and the Israeli Science Foundation.  The Minerva Foundation provided
 postdoctoral fellowships to SK and DK. We are grateful to H. Primack
 and H. Schanz for the spectra of the $2d$ and $3d$ Sinai billiards.
 We thank H. Primack, for performing some critical checks, and for
 many comments and suggestions. We are indebted to F. Haake, for
 reporting about his results prior to their publication, and to
 J. Keating, and Z. Rudnik for stimulating discussions on the
 autocorrelation function of the Riemann $\zeta$ function.


\newpage
\begin{thebibliography}{99}
  \bibliographystyle{unsrt}
  
\bibitem {BohigasLH} O. Bohigias, M.J. Giannoni, and C. Schmit, {\em
    Phys. Rev. Lett.} {\bf 52} (1984) 1.  \bibitem {USLH} U.
  Smilansky, {\em Proceedings of the 1989 Les Houches Summer School in
    ``Chaos and Quantum Physics''} Ed. M.J. Giannoni and J.
  Zinn-Justin (1991) 371.  \bibitem {BerryA400} M.V. Berry, {\em Proc.
    Royal Soc. Lond.} {\bf A 400} (1985) 229.  \bibitem {AA} A.V.
  Andreev and B.L. Altshuler, {\em Phys. Rev. Lett.} {\bf 75} (1995)
  902.  \bibitem {AAA} O. Agam, B.L. Altshuler, and A.V. Andreev, {\em
    Phys. Rev. Lett.} {\bf 75} (1995) 4389.  \bibitem {Bogolkeating}
  E.B. Bogomolny and J. Keating {\em Phys. Rev. Lett.} {\bf 77} (1996)
  1472.  \bibitem {haake} F. Haake, M. Kus, H.-J. Sommers, H.
  Schomerus, and K. Zyckowski {\em J. Phys. A} {\bf 29} (1996) 3641.
\bibitem {Bogomolny} E.B. Bogomolny {\em Nonlinearity} {\bf 5} (1992)
  805.  \bibitem {DoronUS} E. Doron and U. Smilansky, {\em
    Nonlinearity} {\bf 5} (1992) 1055.  \bibitem {KeatingBerry} J.
  Keating and M.V. Berry {\em Proc. Royal Soc. Lond.} {\bf A 437}
  (1992) 151.  \bibitem {Keatingzeta} J. Keating, {\em in
    preparation}.  \bibitem {Dyson} F.J. Dyson, {\em J. Math. Phys. }
  {\bf 3} (1962) 140.  \bibitem {AS} A.V. Andreev and B.D. Simons,
  {\em Phys. Rev. Lett.} {\bf 75} (1995) 2304.  \bibitem{altland} A.
  Altland, S. Iida, K. B. Efetov, {\em J. Phys. A} {\bf 26} (1993)
  3545.  \bibitem{efetov} K. B.  Efetov, {\em Adv. Phys.} {\bf 32}
  (1983) 53.  \bibitem{Metropolis} N. Metropolis, A. Rosenbluth, M.
  Rosenbluth, A. Teller and M. Teller, {\em J. Chem. Phys.} {\bf 21}
  (1953) 1081.  \bibitem {schanz} H. Schanz and U. Smilansky {\em
    Chaos, Solitons and Fractals }{\bf 5} (1995) 1289.  \bibitem
  {primack} H. Primack and U. Smilansky, {\em Phys. Rev. Lett.} {\bf
    74} (1995) 4831.
\bibitem{Plemelj} J. Plemelj, {\em Monat. Math. Phys.} {\bf 15} (1909)
  93.  \bibitem {hannay} H. J. Hannay and A. M. Ozorio de Almeida,
  {\em J. Phys} {\bf A17} (1987) 3429

\end{thebibliography}
\end{document}